\documentclass{article}

\usepackage[hidelinks]{hyperref}

\usepackage{graphicx}
\usepackage{color}
\usepackage{subcaption}
\usepackage{amssymb}
\usepackage{amsthm}
\usepackage{amsmath}

\newtheorem{theorem}{Theorem}

\newtheorem{lemma}{Lemma}

\title{\textsc{MaxCut} on Permutation Graphs is NP-complete}

\author{
    Celina M. H. de Figueiredo$^*$, Alexsander A. de Melo$^{**}$, \\ Fabiano S. Oliveira$^\dagger$, Ana Silva$^\ddagger$ \\ \\
$^{*,**}$Federal University of Rio de Janeiro, Rio de Janeiro, Brazil \\
\texttt{\{celina,aamelo\}@cos.ufrj.br} \\
$^\dagger$Rio de Janeiro State University, Rio de Janeiro, Brazil \\
\texttt{fabiano.oliveira@ime.uerj.br}
\\
$^\ddagger$Federal University of Cear\'{a}, Cear\'{a}, Brazil \\ Universit\'a degli Studi di Firenze, Italy\\
\texttt{anasilva@mat.ufc.br}\\
}

\date{}

\begin{document}

\maketitle

\begin{abstract}
In this paper, we prove that the \textsc{MaxCut} problem is NP-complete on permutation graphs, settling a long-standing open problem that appeared in the 1985 column of the  {\it Ongoing Guide to NP-completeness} by David S. Johnson. 
\end{abstract}

\section{Introduction}
A \emph{cut} is a partition of the vertex set of a graph into two disjoint parts, and the \emph{maximum cut problem} (denoted by \textsc{MaxCut}, for short) aims to determine a cut with the maximum number of edges for which each endpoint is in a distinct part. 
The decision problem \textsc{MaxCut} is known to be NP-complete since the seventies~\cite{GJS76}, 
and only recently its restriction to interval graphs has been announced to be hard by Adhikary, Bose, Mukherjee, and Roy~\cite{ABMR20}. This settles a long-standing open problem from the {\it Ongoing Guide to NP-completeness} by David S. Johnson~\cite{J85}.

In his column, David S. Johnson presented a two-page summary table, with a column for each of the ten most famous NP-complete graph problems, and a row for each of thirty selected graph class. Among those graph classes, special emphasis was given to 
subclasses of perfect graphs and of intersection graphs having broad algorithmic significance.
The emphasis was on the restrictions themselves and how they affect the complexity of the considered NP-hard problems. The discussion had focus on the particularly fertile domain of graph theory, where the central open problem at that time was the recognition of perfect graphs.

Many important graph classes are defined or can be characterized by a geometric intersection model. 
Two particularly well-studied examples are subclasses of perfect graphs: the classes of interval graphs and of permutation graphs~\cite{FIS85,10.5555/984029,Spinrad:1380920}. 
In their respective models, the intersecting objects are line segments in the plane, with  different restrictions imposed on their positions. 
In interval graphs, each line segment must have its endpoints on a single line, while in permutation graphs, their endpoints must lie on two distinct parallel lines.

Besides selecting the recognition of perfect graphs as the famous open problem, in his column, David S. Johnson selected only two others as open and may well be hard problems: Hamiltonian circuit restricted to permutation graphs and edge-coloring restricted to planar graphs. 
Today, we know that recognition of perfect graphs and Hamiltonian circuit restricted to permutation graphs can both be solved in polynomial time. On the other hand, edge-coloring restricted to planar graphs remains a challenging open problem. Please, refer to~\cite{FMSSarxiv} for an updated summary table.
Surprisingly, after 35 years, the only new resolved entry for permutation graphs is Hamiltonian circuit.

The present paper settles a long-standing open problem proposed by Johnson, by providing the first entry of Johnson's table for permutation graphs resolved as NP-complete.
\begin{theorem}\label{theo:maxcut_permutation}
 	{\sc MaxCut} is NP-complete on permutation graphs.
 \end{theorem}
Our proof is based on Adhikary et al.'s construction used to prove the NP-completeness of {\sc MaxCut} on interval graphs~\cite{ABMR20}.
It is interesting to notice that, among the problems selected by Johnson, \textsc{MaxCut} is the only one  classified as NP-complete  for interval graphs and for permutation graphs. 
Despite that, the interval graph constructed by Adhikary et al. is not a permutation graph, and our constructed permutation graph is not an interval graph. 
Thus, we leave as an open question the complexity of \textsc{MaxCut} on permutation interval graphs.

Our paper is organized as follows. In Section~\ref{sec:preliminaries}, we present the basic concepts and notations. In Section~\ref{sec:grained}, we present the main gadget in the reduction of Adhikary et al.~\cite{ABMR20}, which also plays an important role in our reduction. In Section~\ref{sec:Adhikary}, we present the construction of Adhikary et al.~\cite{ABMR20} and show that it does not lead to a permutation graph. The presentation of their construction is also useful in Section~\ref{sec:proof}, where we finally present the proof of Theorem~\ref{theo:maxcut_permutation}.
In Section~\ref{sec:conclusion}, we prove that our constructed permutation graph is not an interval graph, and propose the complexity of \textsc{MaxCut} on permutation interval graphs as an open problem.

\subsection{Preliminaries}\label{sec:preliminaries}
In this work, all graphs considered are simple. 
For missing definitions and notation of graph theory, we refer to~\cite{BM2008}. 

Let $G$ be a graph. We say that a subset $K\subseteq V(G)$ is a \emph{clique} if every two distinct vertices in $K$ are adjacent, and that a subset $S\subseteq V(G)$ is a \emph{stable set} if no two vertices in $S$ are adjacent. 
Let $X$ and $Y$ be two disjoint subsets of $V(G)$. We say that $X$ is \emph{complete} to $Y$ if every vertex in $X$ is adjacent to every vertex in $Y$, and that $X$ is \emph{anti-complete} to $Y$ if there are no edges between $X$ and $Y$. 
We let $E_G(X,Y)$ be the subset of $E(G)$ with an endpoint in $X$ and the other endpoint in $Y$. 
A \emph{cut} of $G$ is a partition of $V(G)$ into two parts $A, B \subseteq V(G)$, denoted by $[A, B]$; 
the edge set $E_G(A,B)$ is called the \emph{cut-set} of $G$ associated with $[A,B]$.  
For each two vertices $u,v \in V(G)$, we say that $u$ and $v$ \emph{are in a same part of $[A,B]$} if either $\{u,v\} \subseteq A$ or $\{u,v\} \subseteq B$; otherwise, we say that $u$ and $v$ \emph{are in opposite parts of $[A,B]$}. 
Denote by $\mathsf{mc}(G)$ the maximum size of a cut-set of $G$. 
The \textsc{MaxCut} problem has as input a graph $G$ and a positive integer $k$, and it asks whether $\mathsf{mc}(G) \geq k$.

Let $\pi$ and $\pi'$ be two permutations of a same set, say $V$.
A graph $G$ is called the \emph{intersection graph related to $\{\pi,\pi'\}$} if $V(G) = V$ and, for each two vertices $u, v \in V(G)$, $uv\in E(G)$ if and only if $u <_{\pi} v$ and $v <_{\pi'} u$. 
In this case, we also say that $\{\pi, \pi'\}$ is a \emph{permutation model} of $G$. 
A graph is a \emph{permutation graph} if  it is the intersection graph related to a permutation model.

Given two permutations $\pi$ and $\gamma$ of disjoint subsets $X$ and $Y$, respectively, we write $\pi\gamma$ to denote the permutation of $X\cup Y$ given by the \emph{concatenation} of $\pi$ with $\gamma$. 
Also, we write $\overleftarrow{\pi}$ to denote the reverse of the permutation $\pi$, that is, if $\pi = (v_1,\ldots,v_i)$, then $\overleftarrow{\pi} = (v_i,\ldots,v_1)$. 
In order to simplify the notation, given a set $Z$, we sometimes use the same symbol, $Z$, to denote also a chosen permutation of the elements of $Z$; in such cases, $\overleftarrow{Z}$ represents the reverse of the chosen permutation for $Z$.

An \emph{interval model} is a finite multiset ${\cal M}$ of closed intervals of the real line. 
Let $G$ be a graph and ${\cal M}$ be an interval model. 
An \emph{${\cal M}$-representation} of $G$ is a bijection $\phi \colon V(G) \rightarrow {\cal M}$ such that, for every two distinct vertices $u, v \in V(G)$, we have that $uv \in E(G)$ if and only if $\phi(u) \cap \phi(v) \neq \emptyset$. 
If such an ${\cal M}$-representation exists, we say that ${\cal M}$ is an \emph{interval model of $G$} and that $G$ 
is an \emph{interval graph}.

We write $i\in [n]$ to mean $i\in \{1, \ldots, n\}$.

\section{Grained gadget}\label{sec:grained}

In this section, we present the notion of \emph{grained gadgets}, which was defined in~\cite{FMOS21} as a generalization of the so-called \emph{$V$-gadgets} and \emph{$E$-gadgets}, these latter introduced by Adhikary et al.~\cite{ABMR20} in order to prove the NP-completeness of {\sc MaxCut} on interval graphs.

\begin{figure}[ht]\centering
	\includegraphics[scale=3.5]{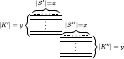}
	\caption{Interval representation of an $(x,y)$-grained gadget~c.f.\cite{FMOS21}.}\label{fig:interval_grained}
\end{figure}

\begin{figure}[ht]\centering\captionsetup[subfigure]{justification=centering}
	\includegraphics[scale = 1.0]{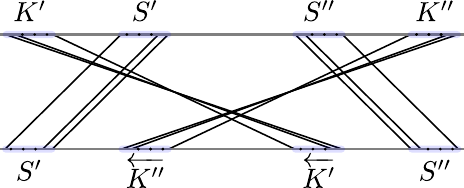}
	\caption{A permutation model of a grained gadget.}\label{fig:grained_permutations}
\end{figure}

Let $x$ and $y$ be positive integers.
An \emph{$(x, y)$-grained gadget} is a split graph $H$ formed by a clique $K'\cup K''$ of size $2y$ and a stable set $S'\cup S''$ of size $2x$ with $K'$ being complete to $S'$, $K''$ being complete to $S''$, and satisfying $|K'| = |K''| = y$ and $|S'| = |S''| = x$. 
Figure~\ref{fig:interval_grained} depicts an interval representation of an $(x,y)$-grained gadget. 
One can readily verify that the intersection graph 
related to the pair of permutations $\{K'S'S''K'', S'\overleftarrow{K''}\overleftarrow{K'}S''\}$ (see Figure~\ref{fig:grained_permutations}) is an $(x,y)$-grained gadget. 
Thus, grained gadgets are interval graphs and permutation graphs.

\begin{figure}[ht]\centering
	\begin{subfigure}[t]{0.19\textwidth}\centering
		\includegraphics[scale = 0.8]{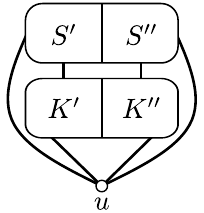}
		\caption{Covering intersection}\label{subfig:covering}
	\end{subfigure}
	\begin{subfigure}[t]{0.38\textwidth}\centering
	    \includegraphics[scale = 0.8]{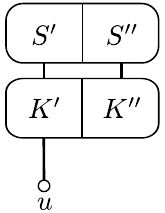}
	    \hspace{2.0ex}
	    \includegraphics[scale = 0.8]{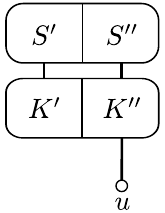}
		\caption{Weak intersection}\label{subfig:weakly}
	\end{subfigure}
\begin{subfigure}[t]{0.38\textwidth}\centering
	    \includegraphics[scale = 0.8]{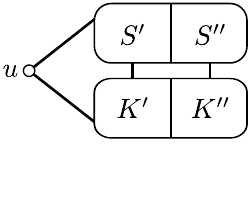}
	    \hspace{2.0ex}
	    \includegraphics[scale = 0.8]{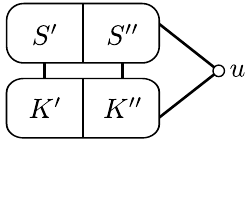}
		\caption{Strong intersection}\label{subfig:strongly}
	\end{subfigure}

	\caption{Vertex $u \in V(G)\setminus V(H)$
	(\protect\subref{subfig:covering}) covering $H$, (\protect\subref{subfig:weakly}) weakly intersecting $H$, and (\protect\subref{subfig:strongly}) strongly intersecting $H$. The set $K'\cup K''$ is a clique and  the set $S'\cup S''$ is a stable set.  A line between sets, or between $u$ and some set, means that all the edges occur. 
	}

\label{fig:intersection_types}
\end{figure}

Let $H$ be an $(x,y)$-grained gadget and $G$ be a supergraph of $H$. 
For each vertex $u \in V(G)\setminus V(H)$, we say that (see Figure~\ref{fig:intersection_types}): $u$ \emph{covers} $H$ if $V(H) \subseteq N_{G}(u)$; $u$ \emph{weakly intersects} $H$ if either $N_{G}(u) \cap V(H) = K'$ or $N_{G}(u) \cap V(H) = K''$; and that $u$ \emph{strongly intersects} $H$ if either $N_{G}(u) \cap V(H) = K' \cup S'$ or $N_{G}(u) \cap V(H) = K'' \cup S''$.
Moreover, we say that $G$ \emph{respects the structure} of $H$ if, for each vertex $u \in V(G)\setminus V(H)$, either $N_{G}(u) \cap V(H) = \emptyset$ or $u$ satisfies one of the previous conditions.

The next lemma establishes the key property of grained gadgets with respect to the {\sc MaxCut} problem. 
Intuitively, it states that, for suitable values of $x$ and $y$, if $G$ is a supergraph that respects the structure of an $(x,y)$-grained gadget, then, in any maximum cut $[A, B]$ of $G$, the vertices belonging to $K' \cup S''$ are placed in a same part of $[A, B]$, opposite to the part containing the vertices belonging to $K'' \cup S'$.

\begin{lemma}[\cite{FMOS20}]Let $x$ and $y$ be positive integers, $H$ be an $(x,y)$-grained gadget and $G$ be a supergraph that respects the structure of $H$. 
	Also, let $[A, B]$ be a maximum cut of $G$, $t$ be the number of vertices in $V(G)\setminus V(H)$ adjacent to some vertex of $H$, $\ell$ be the number of vertices of $G$ adjacent to some vertex in $S'$, and $r$ be the number of vertices of $G$ adjacent to some vertex in $S''$. 
	If $\ell$ and $r$ are odd, $y > 2t$ and $x > t + 2y$, then each of the following holds:
	\begin{enumerate}
		\item $S' \subseteq A$ and $K' \subseteq B$, or vice versa;
		\item $S'' \subseteq A$ and $K'' \subseteq B$, or vice versa;
		\item $K' \subseteq A$ and $K'' \subseteq B$, or vice versa.
	\end{enumerate}
	\label{lem:grainedGadgets}
\end{lemma}

In the remainder of the text, when a grained gadget $H$ is not clear in the context, we write $S'(H)$, $S''(H)$, $K'(H)$ and $K''(H)$ to denote the stable sets $S'$ and $S''$ and the cliques $K'$ and $K''$ of $H$, respectively.

\section{Adhikary et al.'s reduction}\label{sec:Adhikary}

In this section, we present the construction given by Adhikary et al.~\cite{ABMR20} of an interval graph that proves NP-completeness of {\sc MaxCut} in this class. As we see in Section~\ref{sec:proof}, the general idea behind their construction can also be used to obtain a permutation graph instead. Nevertheless, the question of whether their construction is also permutation might arise. We prove here that this is not the case.

\begin{figure}[ht]
    \includegraphics[scale=1.0]{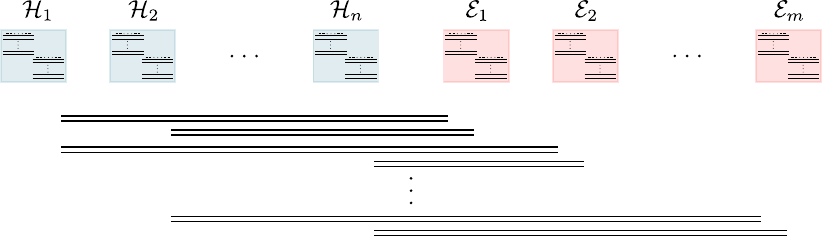}\centering
    \caption{Adhikary et al.'s interval model $\mathcal{M}$, with $e_1=v_1v_2$, $e_2=v_1v_n$, and $e_m=v_2v_n$.}\label{fig:intervalModel}
\end{figure}

Given a cubic graph $G$, let $\pi_{V} = (v_1,v_2,\ldots,v_n)$ and $\pi_{E} = (e_1,e_2,\ldots,e_m)$ be arbitrary orderings of $V(G)$ and $E(G)$, respectively. Define the values: $q = 200n^3 + 1$, $p = 2q + 7n$, $q' = 10n^2 + 1$, and $p' = 2q' + 7n$. An interval graph $G'$ is defined through the construction of one of its interval models ${\cal M}$, described as follows (observe Figure~\ref{fig:intervalModel} to follow the construction):
    \begin{enumerate}
        \item Add to ${\cal M}$ a $(p,q)$-grained gadget ${\cal H}_i$ for each vertex $v_i \in V(G)$. These gadgets should be pairwise disjoint, with ${\cal H}_i$ appearing completely to the left of ${\cal H}_{i+1}$ for every $i\in [n-1]$;
        \item Add to ${\cal M}$ a $(p',q')$-grained gadget ${\cal E}_j$ for each edge $e_j \in E(G)$. Likewise, these gadgets should be pairwise disjoint, with ${\cal E}_j$ appearing completely to the left of ${\cal E}_{j+1}$ for every $j\in [m-1]$. Additionally, ${\cal E}_1$ appears completely to the right of ${\cal H}_n$, without intersecting it;
        \item Finally, for each edge $e_j = v_iv_{i'} \in E(G)$, with $i < i'$, add four intervals $L^1_{i,j}, L^2_{i,j}, L^1_{i',j}, L^2_{i',j}$, called \emph{link}  intervals, such that:
        \begin{itemize}
            \item $L^1_{i,j}$ and $L^2_{i,j}$ (resp. $L^1_{i',j}$ and $L^2_{i',j}$) weakly intersect ${\cal H}_i$ 
            (resp. ${\cal H}_{i'}$) to the right of ${\cal H}_i$ 
            (resp. ${\cal H}_{i'}$);
            \item $L^1_{i,j}$ and $L^2_{i,j}$ (resp. $L^1_{i',j}$ and $L^2_{i',j}$) weakly intersect (resp. strongly intersect) ${\cal E}_j$ to the left  of ${\cal E}_j$.
        \end{itemize}
    \end{enumerate}

\begin{figure}[ht]\centering\captionsetup[subfigure]{justification=centering}
	\begin{subfigure}[t]{0.32\textwidth}\centering
		\includegraphics[scale = 1.0]{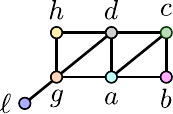}
		\caption{}\label{subfig:forbidden_subgraph}
	\end{subfigure}
	\hspace{2.0ex}
	\begin{subfigure}[t]{0.55\textwidth}\centering
		\includegraphics[scale = 1.0]{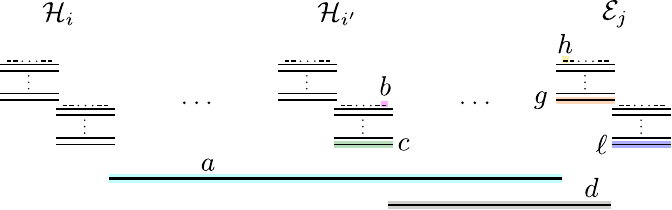}
		\caption{}\label{subfig:forbidden_adhkary}
	\end{subfigure}	
	\caption{\protect(\subref{subfig:forbidden_subgraph})~Forbidden induced subgraph $\overline{X}_{34}$ for comparability graphs~cf.\cite{ISGCI}. \protect(\subref{subfig:forbidden_adhkary})~$\overline{X}_{34}$ as an induced subgraph in Adhikary et al.'s construction.}\label{fig:forbidden}
\end{figure}

As claimed, we show that the constructed graph $G'$ is not a permutation graph. This is because $G'$ contains the graph $\overline{X}_{34}$ depicted in Figure~\ref{subfig:forbidden_subgraph} as an induced subgraph, and such a graph is a forbidden subgraph for comparability graphs cf.~\cite{Gallai1967, ISGCI}, in turn a known superclass of permutation graphs. 
To see that this claim holds, observe Figure~\ref{subfig:forbidden_adhkary}. 
Given an edge $e_j = v_iv_{i'}\in E(G)$, with $i < i'$, it shows the intervals in the grained gadgets of $v_i$, $v_{i'}$ and $e_j$, as well as some link intervals related to $e_j$. 
The adjacencies can be easily checked to be as in the graph of Figure~\ref{subfig:forbidden_subgraph}.

In the next section, we show that a modification of 
Adhikary et al.'s construction gives us the desired permutation graph.

\section{Our reduction}
\label{sec:proof}

Consider a cubic graph $G$, and orders on its vertex and edge sets, $\pi_{V} = (v_1,v_2,\ldots,v_n)$ and $\pi_{E} = (e_1,e_2,\ldots,e_m)$, respectively. The values of $p,q,p',q'$ are not the same as in Section~\ref{sec:Adhikary} and are presented later. 
Again, for each vertex $v_i$, create a $(p,q)$-grained gadget, ${\cal H}_i$, and for each edge $e_j$, create a $(p',q')$-grained gadget ${\cal E}_j$. 
For simplicity, denote the sets $S'({\cal H}_i)$, $S''({\cal H}_i)$, $K'({\cal H}_i)$ and $K''({\cal H}_i)$ by $S'_i,S''_i,K'_i,K''_i$, respectively. Similarly, denote the sets $S'({\cal E}_j)$, $S''({\cal E}_j)$, $K'({\cal E}_j)$ and $K''({\cal E}_j)$ by $S^{'e}_j,S^{''e}_j,K^{'e}_j,K^{''e}_j$, respectively. 

Recall that for each $i \in [n]$, the permutation model of ${\cal H}_i$ consists of the pair of permutations $\{\pi^1_i, \pi^2_i\}$
where $\pi^1_i = K_i' S_i' S_i'' K_i''$ and $\pi^2_i = S'_i  \overleftarrow{K''_i} \overleftarrow{K'_i} S''_i$. Analogously, for each $j \in [m]$, the permutation model of ${\cal E}_j$ consists of the pair of permutations $\{\gamma^1_j, \gamma^2_j\}$
where $\gamma^1_j = K^{'e}_j S^{'e}_j S^{''e}_j K^{''e}_j$ and $\gamma^2_j = S^{'e}_j \overleftarrow{K^{''e}_j} \overleftarrow{K^{'e}_j} S^{''e}_j$.
Now, for each edge $e_j = v_iv_{i'}$, with $i < i'$, add four new vertices $L^1_{i,j}, L^2_{i,j}, L^1_{i',j}, L^2_{i',j}$, called \emph{link}  vertices. In what follows, we modify some of the grained gadget permutations in order to make $L^1_{i,j},L^2_{i,j}$ (resp. $L^1_{i',j},L^2_{i',j}$) weakly intersect ${\cal H}_i$ (resp. ${\cal H}_{i'}$) and strongly intersect (resp. weakly intersect) ${\cal E}_j$.

If $v_i$ is incident to edges $j_1,j_2,j_3$, with $j_1<j_2<j_3$, then modify one of the permutations defining ${\cal H}_i$ to include the link vertices related to $v_i$ as follows:
\[\pi^1_i = K'_iS'_iS''_iC_iK''_i,\]
where $C_i$ denotes the permutation  $L^1_{i,j_1}L^2_{i,j_1}L^1_{i,j_2}L^2_{i,j_2}L^1_{i,j_3}L^2_{i,j_3}$.

Similarly, for each edge 
$e_j = v_iv_{i'}$, $i<i'$, we modify one of the permutations defining ${\cal E}_j$ to include the link vertices related to $e_j$  as follows:

\[\gamma^1_j = K^{'e}_jL^2_{i',j}L^1_{i',j}S^{'e}_jL^2_{i,j}L^1_{i,j}S^{''e}_jK^{''e}_j.\]

\begin{figure}[htb]\centering
    \includegraphics[scale=0.75]{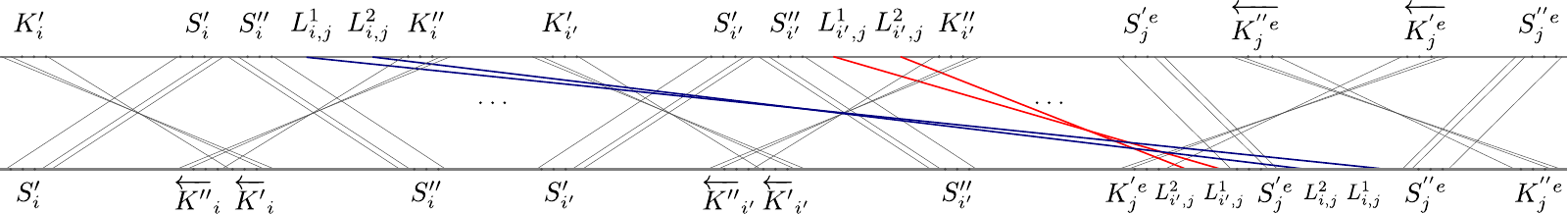}
    \caption{Vertex and edge grained gadgets, and link vertices related to an edge $e_j = v_{i}v_{i'}$, with $i < i'$, in our permutation model $\{\Pi,\Pi'\}$.}\label{fig:permutation_model}
\end{figure}

We do not modify $\pi^2_i$ and $\gamma^2_j$, and keep denoting
by $\pi^2_i$ the permutation  $S'_i\overleftarrow{K''}_i\overleftarrow{K'}_iS''_i$, and by $\gamma^2_j$ the permutation $S^{'e}_j\overleftarrow{K^{''e}}_j\overleftarrow{K^{'e}}_jS^{''e}_j$. 
Finally, let $G'$ be the permutation graph related to $\{\Pi,\Pi'\}$, where:
\[\Pi = \pi^1_1\ldots\pi^1_n\gamma^2_1,\ldots,\gamma^2_m\mbox{, and }\]
\[\Pi' = \pi^2_1\ldots\pi^2_n\gamma^1_1,\ldots,\gamma^1_m.\]
Figure~\ref{fig:permutation_model} illustrates our permutation model $\{\Pi,\Pi'\}$, focusing on the vertex grained gadgets $\mathcal{H}_{i}$ and $\mathcal{H}_{i'}$, the edge grained gadget $\mathcal{E}_{j}$, and the link vertices $L_{i,j}^{1}, L_{i,j}^{2}$ and $L_{i',j}^{1}, L_{i',j}^{2}$ related to an edge $e_j = v_{i}v_{i'}$, with $i < i'$.

We remark that the main difference of our permutation graph from the Adhikary et al.'s interval graph is the fact that, in Adhikary et al.'s interval graph, the link vertices form a clique, whereas, as we show in Section~\ref{sec:conclusion}, some link vertices are not adjacent in our permutation graph. 
Additionally, for an edge $e_j = v_iv_{i'} \in E(G)$, with $i < i'$, the link vertices $L_{i,j}^{1}, L_{i,j}^{2}$ (resp. $L_{i',j}^{1}, 
L_{i' ,j}^{2}$) weakly intersect (resp. strongly intersect) $\mathcal{E}_j$ in Adhikary et al.'s interval graph, whereas in our permutation graph the link vertices $L_{i,j}^{1}, L_{i,j}^{2}$ (resp. $L_{i',j}^{1}, L_{i' ,j}^{2}$) strongly intersect (resp. weakly intersect) $\mathcal{E}_j$.

Before our proof, we make some observations about the constructed graph in order to improve the proof's readability. 
Note that, for each link vertex $L$ and grained gadget $H$, either the relative order between $L$ and $V(H)$ in $\Pi$ is the reverse of their relative order in $\Pi'$, in which case $L$ is complete to $V(H)$, or the relative order is the same in both $\Pi$ and $\Pi'$, in which case $L$ is anti-complete to $V(H)$, or $L$ is related to $H$ according to one of the ways described below.
\begin{itemize}
    \item $L \in \{L^1_{i,j},L^2_{i,j}\}$ and $H = {\cal H}_i$: in this case only the relative orders between $L$ and $K''_i$ are reversed in $\Pi$ and $\Pi'$, i.e., $L$ is complete to $K''_i$ and anti-complete to $V({\cal H}_i)\setminus K''_i$;
    
    \item $L\in \{L^1_{i,j},L^2_{i,j}\}$ and $H = {\cal E}_j$, with $e_j = v_iv_{i'}$, $i<i'$: in this case the relative orders between $L$ and $K^{'e}_j\cup S^{'e}_j$ are reversed in $\Pi$ and $\Pi'$, i.e., $L$ is complete to $K^{'e}_j\cup S^{'e}_j$ and anti-complete to $V({\cal E}_j)\setminus (K^{'e}_j\cup S^{'e}_j)$; or
    
    \item $L\in \{L^1_{i',j},L^2_{i',j}\}$ and $H = {\cal E}_j$, with $e_j = v_iv_{i'}$, $i<i'$: in this case only the relative orders between $L$ and $K^{'e}_j$ are reversed in $\Pi$ and $\Pi'$, i.e., $L$ is complete to $K^{'e}_j$ and anti-complete to $V({\cal E}_j)\setminus K^{'e}_j$.
\end{itemize}

\begin{proof}[\bf Proof of Theorem~\ref{theo:maxcut_permutation}]
Consider the reduction graph $G'$ and its permutation model $\{\Pi,\Pi'\}$ as previously defined.
For each $e_j = v_iv_{i'} \in E(G)$, let $$L(e_j) = \{L^1_{i,j}, L^2_{i,j},L^1_{i',j},L^2_{i',j}\}\text{;}$$ and for each $v_i\in V(G)$, let $$L(v_i) = \{L^1_{i,j},L^2_{i,j}\mid e_j\mbox{ is incident to }v_i\}\text{.}$$ Also, denote the set of link vertices by ${\cal L}$, i.e.  ${\cal L} = \bigcup_{j=1}^m L(e_j)$.

We postpone the assignment of the actual values for $p,q,p',q'$ and, in addition to the conditions necessary for the application of Lemma~\ref{lem:grainedGadgets}, we also ask that $q>6n+p'$ and $p'>2q'>9n^2$.

In what follows, we prove that there exists a bijective relation $f$ between the maximum cuts of the input graph $G$ and the maximum cuts of our permutation graph $G'$. Then, we prove that, for each maximum cut $[X,Y]$ of $G$, $$|E_G(X,Y)|\ge k \;\text{ if and only if }\; |E_{G'}(A,B)|\ge \phi(n,m,k)\text{,}$$ where $[A,B] = f(X,Y)$ and $\phi$ is a well-defined function. Theorem~\ref{theo:maxcut_permutation} immediately follows.

Let $[A,B]$ be a maximum cut of $G'$. In order to define $f$, we first prove some properties relating the partitioning of vertex and edge grained gadgets of $G'$ in $[A,B]$ with the partitioning of the link vertices of $G'$ in $[A,B]$.
More specifically, we prove that the two following properties hold:
\begin{enumerate}
    \item For each vertex $v_i \in V(G)$, if $K''_{i} \subseteq A$, then $\{L_{i,j}^{1}, L_{i,j}^{2}\} \subseteq B$ for each edge $e_j \in E(G)$, with $e_j = v_iv_{i'}$ and $i < i'$;\label{item:first_property}
    \item For each edge $e_j \in E(G)$, with $e_j = v_iv_{i'}$ and $i < i'$, if $\{L_{i,j}^{1}, L_{i,j}^{2}\} \subseteq B$, then $S^{'e}_{j} \subseteq A$.\label{item:second_property}
\end{enumerate}

\emph{Proof of Property \ref{item:first_property}}. Let $v_i\in V(G)$ and suppose that $K''_{i} \subseteq A$. 
For the sake of contradiction, suppose that there exists a link vertex $L\in L(v_i)\cap A$. 
Then, let $[A',B']$ be the cut obtained from $[A,B]$ by setting $A' = A\setminus\{L\}$ and $B' = B\cup\{L\}$. 
Observe that there is a loss of at most $\lvert{\cal L}\rvert+\max\{p',q'\} = \lvert{\cal L}\rvert+p'$ edges between $L$ and ${\cal L}$, and between $L$ and the vertices of the edge grained gadget related to $L$, say $\mathcal{E}_j$, since $K^{'e}_{j}$ and $S^{'e}_{j}$ are always in opposite parts of the cut. On the other hand we gain all the edges between $L$ and the vertices in $K''_i$. Therefore, we get an increase of the cut-set of at least $q$ edges, and a decrease of less than $|{\cal L}| + \max\{p',q'\} = 6n+p'$ edges. 
It follows from the hypothesis $q > 6n + p'$ that $|E_{G'}(A',B')|$ is bigger than $|E_{G'}(A,B)|$, contradicting the maximality of $[A,B]$.

\emph{Proof of Property \ref{item:second_property}}. Consider an edge $e_j\in E(G)$, with $e_j=v_iv_{i'}$ and $i<i'$, and suppose that $\{L_{i,j}^{1}, L_{i,j}^{2}\} \subseteq B$. 
Observe that, because the relative orders among the edge and vertex grained gadgets themselves are the same in $\Pi$ and $\Pi'$, there are no edges between ${\cal E}_j$ and any other grained gadgets of $G'$, i.e., the only vertices outside of ${\cal E}_j$ that can be adjacent to the vertices of ${\cal E}_j$ are those in ${\cal L}$. 
Moreover, Lemma~\ref{lem:grainedGadgets} tells us that the vertices belonging to $K^{'e}_{j} \cup S^{''e}_{j}$ are placed in a same part of $[A,B]$, opposite to the part containing the vertices belonging to $K^{''e}_{j} \cup S^{'e}_{j}$.
More formally, either $K^{'e}_{j}\cup S^{''e}_{j}\subseteq B$ and $K^{''e}_{j}\cup S^{'e}_{j}\subseteq A$, or $K^{'e}_{j}\cup S^{''e}_{j}\subseteq A$ and $K^{''e}_{j}\cup S^{'e}_{j}\subseteq B$.
As a result, switching the vertices of ${\cal E}_j$ of part of the cut does not change, and therefore cannot decrease, the number of edges between the vertices of ${\cal E}_j$ and the vertices belonging to ${\cal L}\setminus L(e_j)$ in the cut-set. 
Consequently, if $S^{'e}_{j} \subseteq A$, then we obtain that there are at least $2p'$ edges in the cut-set that are incident to vertices of $\mathcal{E}_j$; these are the edges between $L^1_{i,j},L^2_{i,j}$ and the vertices belonging to $S^{'e}_j$.
On the other hand, if $S^{'e}_{j} \subseteq B$, then we obtain that there are at most $4q'$ edges in the cut-set that are incident to vertices of $\mathcal{E}_j$; these are the edges between the vertices belonging to $L(e_j)$ and the vertices belonging to $K^{'e}_j$.
Therefore, since $p'>2q'$, we get that $S^{'e}_{j} \subseteq A$ as we wanted to prove. 

\vspace{2.0ex}

We are now ready to prove the existence of the bijective relation $f$. 
For each maximum cut $[X,Y]$ of $G$, let $f(X,Y)$ be the cut $[A,B]$ of $G'$ defined as follows:
\begin{itemize}
    \item For each vertex $v_i\in V(G)$, if $v_i\in X$, then add $K'_i\cup S''_i\cup L(v_i)$ to $A$ and $K''_i\cup S'_i$ to $B$; do the opposite otherwise. 
    \item For each $e_j\in E(G)$, with $e_{j} = v_iv_{i'}$ and $i<i'$, if $L^1_{i,j}\in A$, then add $K^{'e}_j\cup S^{''e}_j$ to $A$ and $K^{''e}_j\cup S^{'e}_j$ to $B$; and do the opposite otherwise. 
\end{itemize}
Based on Properties~\ref{item:first_property} and~\ref{item:second_property}, one can readily verify that $f$ is well-defined and is a bijective relation, as desired.

\vspace{2.0ex}

Now, we count the number of edges in $E_{G'}(A,B)$ as a function of $n$, $m$, $p$, $q$, $p'$, $q'$ and of the size of the cut-set $E_G(X,Y)$. 
First, consider $v_i\in V(G)$. By construction, we know that there are $2pq+q^2$ edges in the cut-set between the vertices of ${\cal H}_i$. Additionally, there are exactly $6$ link vertices weakly intersecting ${\cal H}_i$, while all other link vertices are either complete or anti-complete to $V({\cal H}_i)$. Observe also that the number of link vertices complete to $V({\cal H}_i)$ is exactly equal to $6(i-1)$; these are the link vertices related to $\{v_1,\ldots,v_{i-1}\}$. This gives us a total of $6[q+(i-1)(p+q)]$ edges between the vertices of ${\cal H}_i$ and the vertices belonging to ${\cal L}$ in the cut-set. 
Summing up these values for every $v_{i} \in V(G)$, we get a total of 
\[\alpha_1 = n[2pq+q^2+6q] + 6\sum_{i=1}^{n}\left((i-1)(p+q)\right) = n[2pq+q^2 + 6q + 3(p+q)(n-1)]\] 
edges in the cut-set $E_{G'}(A,B)$ incident to vertex grained gadgets. 
Now, let $e_j \in E(G)$, with $e_j = v_iv_{i'}$ and $i<i'$. 
By construction, we know that there are $2p'q'+(q')^2$ edges of the cut-set between vertices of ${\cal E}_j$, and $2p'$ edges of the cut-set between $L^1_{i,j},L^2_{i,j}$ and the vertices of ${\cal E}_j$. 
Additionally, note that there are exactly $4(m-j)$ link vertices that cover and, therefore, are complete to ${\cal E}_j$; these are the link vertices related to $\{e_{j+1}, \ldots, e_{m}\}$.
This gives us a total of $4(m-j)(p'+q')$ edges between the vertices of ${\cal E}_j$ and the vertices belonging to ${\cal L}\setminus L(e_j)$.
Finally, suppose without loss of generality that $L^1_{i,j}\in A$ (the count is analogous if it is in $B$).
If $v_{i'}\in X$, then we know that $\{L^1_{i',j},L^2_{i',j}\}\subseteq A$ and hence there are no edges in the cut-set between vertices $L^1_{i',j},L^2_{i',j}$ and the vertices of ${\cal E}_j$. 
Otherwise, observe that it follows that $e_j\in E_{G}(X,Y)$ and $\{L^1_{i',j},L^2_{i',j}\}\subseteq B$, and hence we get additional $2q'$ edges in the cut-set; these additional edges are between the link vertices $L^1_{i',j},L^2_{i',j}$ and the vertices belonging to $K^{'e}_{j}$. 
Summing up these values for every $e_j \in E(G)$, we get a total of $\alpha_2+2q'|E(A,B)|$ edges in the cut-set $E_{G'}(A,B)$ incident to edge grained gadgets, where 
\[\begin{array}{rl}\alpha_2 & =  m[2p'q'+(q')^2+2p'] + \sum_{j=1}^m \left(4(m-j)(p'+q')\right)\\
& = m[2p'q'+(q')^2 + 2p' + 2(p'+q')(m-1)].\end{array}\]

Finally, observe that there are at most $|A\cap {\cal L}|\cdot |B\cap {\cal L}|$ edges of the cut-set between link vertices. Note also that $|A\cap {\cal L}| = 6|X|$ since each vertex in $X$ is related to 6 link vertices, which are all placed in $A$. 
Similarly, we have $|B\cap {\cal L}| = 6|Y|$.
This gives us at most $36|X|\cdot|Y|\le 9n^2$ edges in the cut-set between link vertices. Putting everything together, we get:

\[\alpha_1+\alpha_2+2q'|E_{G}(X,Y)| \le |E_{G'}(A,B)|\le \alpha_1+\alpha_2+2q'|E_{G}(X,Y)|+9n^2.\]

By setting $\phi(n,m,k)$ to $\alpha_1+\alpha_2+2q'k$, and knowing that $p,q,p',q'$ will be chosen as functions of $n$ and $m$, we want to prove, as stated in the beginning, that $|E_{G}(X,Y)|\ge k$ if and only if $|E_{G'}(A,B)|\ge \phi(n,m,k)$. If $|E_{G}(X,Y)|\ge k$, then the first inequality gives us that $|E_{G'}(X,Y)|\ge \alpha_1+\alpha_2+2q'k = \phi(n,m,k)$. On the other hand, if $|E_{G'}(A,B)| \ge \phi(n,m,k) = \alpha_1+\alpha_2+2q'k$, then the second inequality gives us that $|E_{G}(X,Y)|\ge k - 9n^2/2q'$. Because we assume that $2q'>9n^2$, it follows that $k - 9n^2/2q' > k-1$ and hence $|E_{G}(X,Y)|\ge k$.

It only remains to set the values of $p,q,p',q'$. Observe that:
\begin{itemize}
    \item For every grained gadget $H$, the total number of vertices in $V(G')\setminus H$ adjacent to $H$ is at most $6n$ (these are exactly the link vertices).
    \item For every vertex grained gadget $H$, the total number of vertices adjacent to some vertex $u\in S'(H)$ is exactly $q+2h$, for some positive integer $h$ (this is because the number of link vertices adjacent to the vertices in $S'(H)$ is always even). The same holds for the number of vertices adjacent to the vertices in $S''(H)$. We then get that the parity of $\ell$ and $r$ in the conditions of Lemma~\ref{lem:grainedGadgets} applied to $H$ depends only on the parity of $q$.
    \item Similarly, if $H$ is an edge grained gadget, then the parity of the total number of vertices adjacent to some $u\in S'(H)\cup S''(H)$ is equal to the parity of $q'$.
\end{itemize}

Therefore, the necessary conditions of Lemma~\ref{lem:grainedGadgets} translate to: $q > 6n$ and $q' > 6n$; $p > 2q + 6n$ and $p' > 2q' + 6n$; and $q$ and $q'$ are odd. Additionally, we need to ensure: $q>6n+p'$ and $p'>2q'>9n^2$. Hence, consider:
\begin{itemize}
    \item $q'=5n^2+1$;
    \item $p' = 11n^2+6n$;
    \item $q = 12n^2+12n +1$;
    \item $p = 25n^2 + 30n$.
\end{itemize}
Since $n \geq 4$, one can verify that the values described above satisfy all the required conditions. 
This, therefore, concludes the proof of Theorem~\ref{theo:maxcut_permutation}.
\end{proof}

\section{\textsc{MaxCut} on permutation interval graphs is an open problem}\label{sec:conclusion}
In this paper, we have presented a proof of NP-completeness for the \textsc{MaxCut} problem when constrained to permutation graphs. Surprisingly enough, we found that the main gadget in the reduction recently presented by Adhikary et al.~\cite{ABMR20} for interval graphs is also a permutation graph. Additionally, in Section~\ref{sec:Adhikary}, we have seen that being permutation is not a property that holds for the full
construction of Adhikary et al.~\cite{ABMR20}. On the other hand, since the grained gadgets play an important role in our reduction too, one could wonder whether our construction instead is in the intersection between interval and permutation graphs. The answer to that is no as we argue next.

\begin{figure}[ht]\centering
    \includegraphics[scale=1.0]{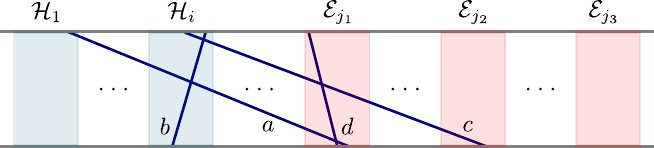}
    \caption{Existence of a $C_4=(a,b,c,d)$ as an induced subgraph in our permutation graph.}\label{fig:c4}
\end{figure}

Let $G$ be a cubic graph, and consider arbitrary orderings of $V(G)$ and $E(G)$, $(v_1,\ldots, v_n)$ and $(e_1,\ldots, e_m)$, respectively. Let $j_1,j_2,j_3$ be the indices of the edges incident to $v_1$, with $j_1 < j_2 < j_3$. Also, let $v_i$ be the other endpoint of $e_{j_2}$. We present a $C_4$ in $G'$, the graph constructed in Section~\ref{sec:proof}; it thus follows that $G'$ is not chordal, and hence also not interval~\cite{10.5555/984029}.
Observe Figure~\ref{fig:c4} to follow our argument. Let $a$ be equal to $L^1_{1,j_1}$, $b$ be any vertex in $K''_i$, $c$ be equal to $L^1_{i,j_2}$, and $d$ be any vertex in $K^{'e}_{j_1}$. Since $j_1 < j_2$ and $1 < i$, we know that the relative order between $a$ and $c$ in $\Pi$ is the same as in $\Pi'$; hence $ac\notin E(G')$. Also, the relative order in $\Pi$ between $a$ and any vertex of ${\cal H}_i$  is reversed in $\Pi'$, the same holds between $c$ and any vertex belonging to $K''_i$; hence $\{ab,bc\}\subseteq E(G')$. Similarly, the relative order between $a$ and any vertex belonging to $K^{'e}_{j_1}$ in $\Pi$ is 
reversed in
$\Pi'$, and the same holds between $c$ and any vertex of ${\cal E}_{j_1}$; hence $\{ad,cd\}\subseteq E(G')$. Finally, since $j_1 < j_2$, the relative order between $b$ and $d$ in $\Pi$ is the same as in $\Pi'$, and therefore $bd\notin E(G')$, thus finishing our argument.

The previous paragraph tells us that for any chosen orderings of $V(G)$ and $E(G)$, the graph constructed in Section~\ref{sec:proof} contains a $C_4$.  
Since it is known that the class of $C_4$-free co-comparability graphs is precisely the class of interval graphs~\cite{Gilmore1964}, and that the class of permutation graphs is equal to the class of comparability co-comparability graphs~\cite{Pnueli1971TransitiveOO}, we get that interval permutation graphs are exactly the class of $C_4$-free permutation graphs.

A good question is whether there is a construction that produces a permutation graph that is also $C_4$-free (and hence interval). 
Up to our knowledge, the largest class in the intersection of permutation and interval graphs for which the complexity is known is the class of the threshold graphs, on which \textsc{MaxCut} is polynomial-time solvable thanks to the algorithm given for cographs, a subclass of permutation graphs that is a superclass of threshold graphs~\cite{bodlaender1994}.

\bibliographystyle{abbrv}

\bibliography{references}

\end{document}